\documentclass[a4paper]{jpconf}
\usepackage{graphicx}
\usepackage{amsmath,amssymb,amsfonts}

\usepackage{setspace}
\usepackage[T1]{fontenc}
\usepackage[latin1]{inputenc}
\usepackage{epsfig}
\usepackage[english]{babel}
\usepackage{color}
\usepackage{dcolumn}
\usepackage{moreverb}

\begin{document}

\title{An infinite family of superintegrable systems from higher order ladder operators and supersymmetry}

\author{Ian Marquette}

\address{Department of Mathematics, University of York, Heslington, York, UK. YO10 5DD}

\ead{im553@york.ac.uk}

\begin{abstract}
We discuss how we can obtain new quantum superintegrable Hamiltonians allowing the separation of variables in Cartesian coordinates with higher order integrals of motion from ladder operators. We also discuss how higher order supersymmetric quantum mechanics can be used to obtain systems with higher order ladder operators and their polynomial Heisenberg algebra. We  present a new family of superintegrable systems involving the fifth Painlev\'e transcendent which possess fourth order ladder operators constructed from second order supersymmetric quantum mechanics. We present the polynomial algebra of this family of superintegrable systems.
\end{abstract}

\section{Introduction}
The most well known superintegrable systems are the Kepler-Coulomb system and the harmonic oscillator [1]. Superintegrable systems with second order integrals of motion in two-dimensional [2] and three dimensional Euclidean spaces [3,4] were classified. However, a systematic search for superintegrable systems with third order integrals of motion in classical and quantum mechanics is more recent and was started in Ref.5. Classical and quantum systems with a second and a third order integrals of motion allowing separation of variables in Cartesian coordinates were obtained by S.Gravel [6] and five of the 14 quantum systems were written in terms of the first, second and fourth Painlev\'e transcendents [7]. These quantum systems were studied [8-10] from the point of view of supersymmetric quantum mechanics and their polynomial algebras. For a review of two-dimensional superintegrable systems we refer the reader to the following paper [11]. In recent years, many articles were devoted to integrable and superintegrable systems with higher order integrals of motion [12-31]. 

A relation between first and second order ladder operators and integrals of motion of superintegrable systems were acknowledged by many authors [1-4,10,15,16,32-34]. A relation between first order supercharges [35] and integrals of motion were also discussed [36-39]. Recently, we pointed out [8-10,12,13] how superintegrability with higher order integrals, higher order ladder operators and higher order supersymmetric quantum mechanics [40-44] are related. We discussed [10] how from known one-dimensional Hamiltonians that possess ladder operators we can form  multi-dimensional superintegrable systems, their integrals and their polynomial algebra. We constructed a new infinite family of superintegrable systems involving the fourth Painlev\'e transcendent from third order ladder operators. We presented a classical analog of this construction and obtained an infinite family of classical superintegrable using third order ladder operators [12]. The factorization of Hamiltonians in supersymmetric quantum mechanics is not necessarily unique and can be used as a tool for obtaining from known systems new superintegrable systems with higher order integrals of motion [13].

The purpose of this paper is to discuss how from systems allowing two intertwining relations with second order supercharges [44], which are related to fourth order ladder operators, we can generate an infinite family of superintegrable systems involving the fifth Painlev\'e transcendent. In Section 2, we present a method to construct superintegrable systems introduced in Ref.10. In Section 3, we discuss the relation between supersymmetric quantum mechanics and ladder operators. In Section 4, we recall results of Ref.44 and using these results we present explicitly the system with fourth order ladder operators, its ladder operators and zero modes in terms of the fifth Painlev\'e transcendent. In Section 5, using results of Section 2 and Section 4, we obtain a new family of superintegrable systems and we present their integrals of motion and the polynomial algebra generated by these integrals.

\section{Superintegrability and ladder operators}
Let us consider the following two-dimensional Hamiltonian allowing separation of variables in Cartesian coordinates:
\begin{equation}
H=H_{1}+H_{2}=\frac{P_{1}^{2}}{2}+\frac{P_{2}^{2}}{2}+V_{1}(x_{1})+V_{2}(x_{2}).
\end{equation}
We suppose the existence of polynomial ladder operators ($\omega_{i}$ are constants and $Q_{i}$ polynomials of order $k_{i}$) :
\begin{equation}
[H_{i},A_{i}^{\dagger}]=\omega_{i}A_{i}^{\dagger},\quad [H_{i},A_{i}]=-\omega_{i}A_{i},
\end{equation}
\begin{equation}
A_{i}A_{i}^{\dagger}=Q_{i}(H_{i}+\omega_{i}),\quad A_{i}^{\dagger}A_{i}=Q_{i}(H_{i}),\quad i=1,2. 
\end{equation}
The polynomial Heisenberg algebra generated by $\{H_{i},A_{i},A_{i}^{\dagger}\}$ provides information on the spectrum of $H_{i}$. The annihilation operator can allow at most $k_{i}$ zero modes (i.e. a state such $A_{i}\psi=0$). In each axis we can have at most $k_{i}$ infinite ladders by acting iteratively with the creation operator. The creation operator can also allow zero modes and we can have finite ladders.

From these operators, we construct the following integrals of motion (of order 2, $k_{1}n_{1}+k_{2}n_{2}$ and $k_{1}n_{1}+k_{2}n_{2}$) with $n_{1}\omega_{1}=n_{2}\omega_{2}=\omega$, $n_{1}$,$n_{2}$ $\in \mathbb{N}^{*}$ :
\begin{equation}
K=\frac{1}{2\omega}(H_{1}-H_{2}),\quad I_{-}=A_{1}^{n_{1}}A_{2}^{\dagger n_{2}},\quad I_{+}=A_{1}^{\dagger n_{1}}A_{2}^{n_{2}}.
\end{equation}
The systems is thus superintegrable. We can also consider the integrals $I_{1}=I_{-}-I_{+}$ and $I_{2}=I_{-}+I_{+}$. These results can be used to study known superintegrable systems and also construct new ones. Moreover, this construction can be extended for N-dimensional quantum systems and also in context of classical systems [12]. These integrals of motion generate a polynomial algebra [8-12,45-48] of order ($k_{1}n_{1}+k_{2}n_{2}-1$)
\begin{equation}
[K,I_{\pm}]=\pm I_{\pm}, \quad [I_{-},I_{+}]=F_{n_{1},n_{2}}(K+1,H)-F_{n_{1},n_{2}}(K,H),
\end{equation}
\begin{equation}
F_{n_{1},n_{2}}(K,H)=\prod_{i=1}^{n_{1}}Q_{1}(\frac{H}{2}+n_{1}\omega_{1}K-(n_{1}-i)\omega_{1})\prod_{j=1}^{n_{2}}Q_{2}(\frac{H}{2}-n_{2}\omega_{2}K+j\omega_{2}).
\end{equation}
The equations (5) and (6) form a deformed $u(2)$ algebra and the finite dimensional representation modules can be found using a realization as a generalized deformed oscillator algebra [49]. The operators $b^{\dagger}=I_{+}$, $b=I_{-}$, $N=K-u$ ( $u$ is a constant) and $\Phi(N,u,H)=F_{n_{1},n_{2}}(K,H)$ satisfy a deformed oscillator algebra :
\begin{equation}
[N,b^{t}]=b^{t} ,\quad [N,b]=-b,\quad b^{t}b=\Phi(N,u,H),\quad bb^{t}=\Phi(N+1,u,H) \quad .
\end{equation}
The function $\Phi(N)$ is called the "structure function". Following [7-10,47,48] we have an energy dependent Fock space of dimension p+1 if
\begin{equation}
\Phi(0,u,E)=0 ,\quad \Phi(p+1,u,E)=0, \quad \Phi(n,u,E) >0, \quad \quad  n=1,2,...,p.
\end{equation}
The Fock space is defined by
\begin{equation}
H|E,n>=E|E,n>,\quad N|E,n>=n|E,n> \quad b|E,0>=0,
\end{equation}
\begin{equation}
b^{t}|n>=\sqrt{\Phi(n+1,u,E)}|E,n+1>,\quad b|n>=\sqrt{\Phi(n,u,E)}|E,n-1>.
\end{equation}
The relations given by equation (8) can be used to obtain the finite dimensional unitary representations and the corresponding degenerate energy spectrum of superintegrable systems [8-11,47,48]. 
\subsection{Systems with ladder operators of first, second and third order}
Infinite families of superintegrable systems in two-dimensional Euclidean space from first [1], second [15] and third [10] order ladder operators were discussed in previous articles (with $n_{1}\omega_{1}=n_{2}\omega_{2}=\omega$, $n_{1}$,$n_{2}$ $\in \mathbb{N}^{*}$):
\begin{equation}
V(x_{1},x_{2})=\frac{1}{2}(\omega_{1}^{2}x_{1}^{2}+\omega_{2}^{2}x_{2}^{2}),
\end{equation}
\begin{equation}
V(x_{1},x_{2})=\frac{1}{2}(\omega_{1}^{2}x_{1}^{2}+\omega_{2}^{2}x_{2}^{2})+\frac{l_{1}}{x_{1}^{2}}+\frac{l_{2}}{x_{2}^{2}},
\end{equation}
\begin{equation}
V(x_{1},x_{2})=\frac{\omega_{1}^{2}}{2}(x_{1}+ \sqrt{\frac{\hbar}{\omega_{1}}}P_{4(1)}(\sqrt{\frac{\omega_{1}}{\hbar}}x_{1}),\alpha_{1},\beta_{1})^{2} + \frac{\epsilon_{1}\hbar\omega_{1}}{2}P_{4(1)}^{'}(\sqrt{\frac{\omega_{1}}{\hbar}}x_{1},\alpha_{1},\beta_{1})+\frac{\hbar\omega_{1}}{3}(\epsilon_{1}-\alpha_{1})
\end{equation}
\[+ \frac{\omega_{2}^{2}}{2}(x_{2}+ \sqrt{\frac{\hbar}{\omega_{2}}}P_{4(2)}(\sqrt{\frac{\omega_{2}}{\hbar}}x_{2},\alpha_{2},\beta_{2}))^{2} +\frac{\epsilon_{2}\hbar\omega_{2}}{2}P_{4(2)}^{'}(\sqrt{\frac{\omega_{2}}{\hbar}}x_{2},\alpha_{2},\beta_{2})+\frac{\hbar\omega_{2}}{3}(\epsilon_{2}-\alpha_{2})\]
\begin{equation}
P_{4(i)}^{''}(z_{i}) = \frac{P_{4(i)}^{'2}(z_{i})}{2P_{4(i)}(z_{i})} + \frac{3}{2}P_{4(i)}^{3}(z_{i}) + 4zP_{4(i)}^{2}(z_{i}) + 2(z_{i}^{2} -
\alpha_{i})P_{4(i)}(z_{i}) +  \frac{\beta_{i}}{P_{4(i)}(z_{i})},
\end{equation}
\[P_{4(i)}(z_{i})=P_{4(i)}(z_{i},\alpha_{i},\beta_{i}),\quad z_{i}=\sqrt{\frac{\omega_{i}}{\hbar}}x_{i}.\]
The integrals of motion, the polynomial algebras , the finite-dimensional unitary repsentations and the corresponding degenerate energy spectrum were given respectively for superintegrable systems with first order ladder operators in Ref. 33 and systems with second and third order ladder operators in Ref. 10.
\section{Supersymmetry and ladder operators}
Let us recall some results of first and second order supersymmetric quantum mechanics (also known as intertwining or factorization method) [35]

Let us consider the following intertwining relation
\begin{equation}
H_{a}L^{\dagger}=L^{\dagger}H_{b},
\end{equation}
with a first order intertwining operator (also called supercharge) and two Hamiltonians of the form 
\begin{equation}
H_{i}=-\frac{\hbar^{2}}{2}\frac{d^{2}}{dx^{2}}+V_{i}(x),\quad L^{\dagger}=\frac{1}{\sqrt{2}}(-\hbar\frac{d}{dx}+\alpha(x)), \quad i=a,b.
\end{equation}
These relations lead to
\begin{equation}
V_{b}=V_{a}-\hbar\alpha',\quad 2\alpha V_{b}-\hbar^{2}\alpha''=2\alpha V_{a}-2\hbar V_{a}'.
\end{equation}
We obtain by putting the first relation in the second and integrating the result
\begin{equation}
\hbar \alpha'+\alpha^{2}=2(V_{a}(x)-\epsilon).
\end{equation}
These equations are equivalent to the factorization method 
\begin{equation}
H_{a}=L^{\dagger}L+\epsilon,\quad H_{b}= LL^{\dagger}+\epsilon.
\end{equation}
If one of these systems is exactly solvable and allows ladder operators, we can obtained the wavefunctions (up to a possible zero mode state), the ladder operators and the energy spectrum of its superpartners using the supercharges. The operators $A_{a}^{\dagger}$ and $A_{b}^{\dagger}$ ($A_{a}$ and $A_{b}$) respectively the raising (lowering) operators of the Hamiltonians $H_{a}$ and $H_{b}$ are related by $A_{a}^{\dagger}=L^{\dagger}A_{b}^{\dagger}L$ ($A_{a}=L^{\dagger}A_{b}L$). Supersymmetric quantum mechanics can thus be used to generate systems with higher order ladder operators.

Let us now consider two Hamiltonians [40-44]
\begin{equation}
L^{\dagger}=\frac{1}{2}(\hbar^{2}\frac{d^{2}}{dx^{2}}-\hbar g(x)\frac{d}{dx}+h(x)).
\end{equation}
We obtain the following three equations
\begin{equation}
V_{b}=V_{a}-\hbar g(x)',
\end{equation}
\begin{equation}
\hbar^{2}\frac{g(x)''}{2}-\hbar h(x)'-g(x)V_{b}=2\hbar V_{a}'-g(x)V_{a},
\end{equation}
\begin{equation}
2h(x)V_{b}-\hbar^{2} h(x)''=2\hbar^{2}V_{0}'' -2\hbar g(x)V_{a}'+2h(x)V_{a}.
\end{equation}
From the equations (21), (22) and (23) we can obtain 
\begin{equation}
V_{a,b}(x)=\hbar^{2}\frac{g(x)''}{4g(x)}-\hbar^{2}\frac{g(x)'^{2}}{8g(x)^{2}}\pm \hbar \frac{g(x)'}{2}+\frac{g(x)^{2}}{8}+\frac{d}{2 \hbar}+\frac{c}{2g(x)^{2}\hbar},
\end{equation}
\begin{equation}
h(x)=-\hbar^{2}\frac{g(x)''}{2g(x)}+\hbar^{2}\frac{g(x)'^{2}}{4g(x)^{2}}-\hbar \frac{g(x)'}{2}+\frac{g(x)^{2}}{4}-\frac{c}{g(x)^{2}\hbar}.
\end{equation}
Moreover, we have
\begin{equation}
LL^{\dagger}=((H_{a}-\frac{d}{2\hbar})^{2}-\frac{c}{4\hbar}),\quad L^{\dagger}L=((H_{b}-\frac{d}{2\hbar})^{2}+\frac{c}{4\hbar}).
\end{equation}
The ladder operators of the Hamiltonians $H_{a}$ and $H_{b}$ are thus related by the second order supercharges.

A first and second order supersymmetry can also be combined [41] to obtain systems with third order ladder operators related  to superintegrable systems obtained by S.Gravel [6,8-10]. If we consider a first $H_{a}L_{1}^{\dagger}= L_{1}^{\dagger}(H_{b}+\hbar\omega)$ and a second order intertwining relations $H_{a}L_{2}^{\dagger}=L_{2}^{\dagger}H_{b}$, we can construct the following third order supercharges $A^{\dagger}=L_{1}^{\dagger}L_{2}$ and $A=L_{2}^{\dagger}L_{1}$. They satisfy respectively the following relations : $[H_{a},A^{\dagger}]=\hbar\omega A^{\dagger}$ and $[H_{a},A]=-\hbar\omega A$. These operators can thus be interpreted as raising and lowering operators. These potentials can be written in terms of the fourth Painlev\'e transcendent. The equation (13) consists in a sum of two such one-dimensional systems.
\section{Systems with fourth order ladder operators}
Let us recall results from Ref.44 and consider two Hamiltonians
\begin{equation}
H_{i}=-\frac{\hbar^{2}}{2}\frac{d^{2}}{dx^{2}}+V_{i}(x),\quad i=a,b,
\end{equation}
intertwined by two second order supercharges $L_{1}$ and $L_{2}$
\begin{equation}
L_{1}^{\dagger}=\frac{1}{2}(\hbar^{2}\frac{d^{2}}{dx^{2}}-\hbar g_{1}(x)\frac{d}{dx}+h_{1}(x)),\quad L_{2}^{\dagger}=\frac{1}{2}(\hbar^{2}\frac{d^{2}}{dx^{2}}-\hbar g_{2}(x)\frac{d}{dx}+h_{2}(x)).
\end{equation}
The interwining equations have the following form
\begin{equation}
H_{b}L_{2}^{\dagger}=L_{2}^{\dagger}H_{a},
\end{equation}
\begin{equation}
H_{a}L_{1}^{\dagger}=L_{1}^{\dagger}(H_{b}+\hbar \omega).
\end{equation}
We construct from these supercharges the following fourth order ladder operators of the Hamiltonian $H_{a}$ and their polynomial Heisenberg algebra
\begin{equation}
A^{\dagger}=L_{1}^{\dagger}L_{2}^{\dagger},\quad A=L_{2}L_{1},
\end{equation}
\begin{equation}
[H_{a},A^{\dagger}]=\hbar\omega A^{\dagger},\quad [H_{a},A]=-\hbar\omega A,
\end{equation}
\begin{equation}
A^{\dagger}A=Q(H_{a})=\prod_{j}^{4}(H_{a}-\epsilon_{j}), \quad AA^{\dagger}=Q(H_{a}+\hbar\omega),
\end{equation}
with $\epsilon_{1,2}=\hbar\omega+\frac{d_{2}}{2\hbar}\pm\sqrt{\frac{c_{2}}{4\hbar}}$, $\epsilon_{3,4}= \frac{d_{1}}{2\hbar}\pm\sqrt{\frac{c_{1}}{4\hbar}}$. The operators ($A^{\dagger}=L_{2}^{\dagger}L_{1}^{\dagger}$ and $A=L_{1}L_{2}$) are ladder operators of the Hamiltonian $H_{2}$.

The intertwining relations (29) and (30) give six equations. Using the following changes of variables $z=\sqrt{\frac{\omega}{\hbar}}x$, $g_{i}=\sqrt{\hbar\omega}\tilde{g_{i}}$,  $h_{i}=\omega \hbar \tilde{h_{i}}$ (with i=1,2), $V_{k}=\hbar\omega \tilde{V_{k}}$ (with k=a,b) and $\epsilon_{j}'= \frac{\epsilon_{j}}{\hbar\omega}$ (with $j=1,2,3,4$) we obtain
\begin{equation}
\frac{\tilde{g_{1}}''}{2\tilde{g_{1}}}-(\frac{\tilde{g_{1}}'}{2\tilde{g_{1}}})^{2}\mp\tilde{g_{1}}'+\frac{\tilde{g_{1}}^{2}}{4}+\frac{(\epsilon_{3}'-\epsilon_{4}')^{2}}{\tilde{g_{1}}^{2}}+\epsilon_{3}'+\epsilon_{4}'=2\tilde{V_{a,b}},
\end{equation}
\begin{equation}
\frac{\tilde{g_{2}}''}{2\tilde{g_{2}}}-(\frac{\tilde{g_{2}}'}{2\tilde{g_{2}}})^{2}\pm\tilde{g_{2}}'+\frac{\tilde{g_{2}}^{2}}{4}+\frac{(\epsilon_{1}'-\epsilon_{2}')^{2}}{\tilde{g_{2}}^{2}}+\epsilon_{1}'+\epsilon_{2}'-1\mp 1=2\tilde{V_{a,b}},
\end{equation}
\begin{equation}
-\frac{\tilde{g_{1}}''}{2\tilde{g_{1}}}+(\frac{\tilde{g_{1}}'}{2\tilde{g_{1}}})^{2}-\frac{\tilde{g_{1}}'}{2}+\frac{\tilde{g_{1}}^{2}}{4}-\frac{(\epsilon_{3}'-\epsilon_{4}')^{2}}{\tilde{g_{1}}^{2}}=h_{1},
\end{equation}
\begin{equation}
-\frac{\tilde{g_{2}}''}{2\tilde{g_{2}}}+(\frac{\tilde{g_{2}}'}{2\tilde{g_{2}}})^{2}-\frac{\tilde{g_{2}}'}{2}+\frac{\tilde{g_{2}}^{2}}{4}-\frac{(\epsilon_{1}'-\epsilon_{2}')^{2}}{\tilde{g_{2}}^{2}}=h_{2},
\end{equation}
\begin{equation}
\tilde{V_{b}}=\tilde{V_{a}}+\tilde{g_{1}}'-1=\tilde{V_{a}}-\tilde{g_{2}}'.
\end{equation}
From the equations (34), (35) and (38), we can obtain the following nonlinear second order differential equation [44] 
\begin{equation}
\tilde{g_{2}}''=\frac{(-2\tilde{g_{2}}+z)}{2\tilde{g_{2}}(-\tilde{g_{2}}+z)}(\tilde{g_{2}}')^{2}+\frac{\tilde{g_{2}}}{z(-\tilde{g_{2}}+z)}\tilde{g_{2}}'++[2z\tilde{g_{2}}(-\tilde{g_{2}}+z)]^{-1}[-2z\tilde{g_{2}}^{5}+(5z^{2}+8\epsilon +4)\tilde{g_{2}}^{4}
\end{equation}
\[-4z(z^{2}+4\epsilon +2)\tilde{g_{2}}^{3}+[z^{4}+4(2\epsilon +1)z^{2}+4(\Delta_{2}^{2}-\Delta_{1}^{2})-1]\tilde{g_{2}}^{2}-4\Delta_{1}^{2}z(-2\tilde{g_{2}}+z)]\]
with ( $\epsilon=\frac{1}{2}(\epsilon_{3}'+\epsilon_{4}') -\frac{1}{2}(\epsilon_{1}'+\epsilon_{2}')$, $\Delta_{1}=\epsilon_{1}'- \epsilon_{2}'$, $\Delta_{2}=\epsilon_{3}' -\epsilon_{4}'$). 

The equation (39) can be transformed into the fifth Painlev\'e transcendent by considering ($\tilde{g_{2}}=\frac{-z}{W-1}$ and $y=z^{2}$)
\begin{equation}
\frac{d^{2}W}{dy^{2}}=(\frac{1}{2W}+\frac{1}{W-1})(\frac{dW}{dy})^{2}-\frac{1}{y}\frac{dW}{dy}+\frac{(W-1)^{2}}{y^{2}}(\frac{aW^{2}+b}{W})+\frac{cW}{y}+\frac{dW(W+1)}{W-1},
\end{equation}
with the following parameters $a=\frac{\Delta_{1}^{2}}{2}$, $b=-\frac{\Delta_{2}^{2}}{2}$, $c=-\epsilon-\frac{1}{2}$, and $d=-\frac{1}{8}$. 
\subsection{Systems in terms of the fifth Painlev\'e transcendent}
Using these results, we present the Hamiltonians $H_{a}$ and $H_{b}$, the supercharges $L_{1}$, $L_{2}$, the ladder operators $L$ and $L^{\dagger}$ and their zero modes in terms of the fifth Painlev\'e transcendent. 

The potential $V(x)=V_{a}(x)-\frac{1}{2}(\epsilon_{1}+\epsilon_{2})$ is given by the following expression
\begin{equation}
V(x)=\frac{\omega^{2}}{8}(1+ \frac{4(P_{5}+P_{5}')^{2}-P_{5}^{2}}{(P_{5}-1)^{2}P_{5}})x^{2}+ \frac{\hbar^{2}}{x^{2}}(a-b-\frac{1}{8}+\frac{b-a P_{5}^{2}}{P_{5}}) -\hbar\omega(1+\frac{(1+2cP_{5})}{2(P_{5}-1)}),
\end{equation}
with ($P_{5}=P_{5}(\frac{\omega}{\hbar}x^{2}, a,b,c,-\frac{1}{8})$). 

The functions $h_{1}$,$h_{2}$,$g_{1}$ and $g_{2}$ in terms of the fifth Painlev\'e transcendent are given by
\begin{equation}
g_{1}=\frac{\omega x}{P_{5}-1}+\sqrt{\frac{\omega}{\hbar}}x,\quad h_{1}=\hbar\omega(\frac{2(4c-P_{5})}{4(-1+P_{5})})
+\hbar^{2}(\frac{-8b+P_{5}(1-8a+8b+8aP_{5})}{4x^{2} P_{5}})
\end{equation}
\[+\omega^{2}(\frac{x^{2}(P_{5}(-1+(-1+P_{5})P_{5})+4P_{5}P_{5}'-4P_{5}'^{2})}{4(-1+P_{5})^{2}P_{5}}),\]
\begin{equation}
g_{2}=-\frac{\omega x}{P_{5}-1},\quad h_{2}=\hbar\omega(\frac{2(1+4cP_{5})}{4(-1+P_{5})})
+\hbar^{2}(\frac{-8b+P_{5}(1-8a+8b+8a P_{5})}{4 x^{2}P_{5}})
\end{equation}
\[+\omega^{2}(\frac{-x^{2}(P_{5}^{2}+P_{5}^{3}+4P_{5}'^{2}+P_{5}(-1+4P_{5}'))}{4(-1+P_{5})^{2} P_{5}}).\]
There are four possible zero modes for the annihilation operator
\begin{equation}
\psi_{1,2}^{(i)} \propto 
(\pm  \sqrt{\frac{a}{2}}(-1+P_{5})-(\frac{1}{2}+c)+ \frac{\frac{\omega}{\hbar}x^{2}(P_{5}+2P_{5}')}
{4 (-1+P_{5})})e^{\int \frac{-1+\frac{\omega}{\hbar}x^{2} \mp \sqrt{8a}(-1+P_{5})^2+P_{5}-2 \frac{\omega}{\hbar}x^{2} P_{5}'}{2 x(-1+P_{5})}dx}, 
\end{equation}
\begin{equation}
\psi_{3,4}\propto e^{\int (\frac{\pm \sqrt{-8b}(-1+P_{5})^2-\frac{\omega}{\hbar}x^{2}(P_{5}^2-2P_{5}')}{2 x(-1+P_{5})P_{5}}+\frac{1}{2x})dx},
\end{equation}
with the corresponding energy levels
\begin{equation}
E_{\psi_{1,2}}=\pm \hbar\omega\sqrt{\frac{a}{2}},\quad E_{\psi_{3,4}}=\hbar\omega(-c-\frac{1}{2}\mp \sqrt{\frac{-b}{2}}),
\end{equation}
and also four possible zero modes for the creation operator
\begin{equation}
\phi_{1,2}\propto e^{ \int \frac{-1-\frac{\omega}{\hbar}x^{2} \pm \sqrt{8a}(-1+P_{5})^2+P_{5}-2 \frac{\omega}{\hbar}x_{i}^{2} P_{5}'}{2 x(-1+P_{5})}dx},\quad 
\end{equation}
\begin{equation}
\phi_{3,4}\propto (\mp  \sqrt{\frac{b}{2}}(-1+P_{5})+(\frac{1}{2}+c)- \frac{\frac{\omega}{\hbar}x^{2}(P_{5}+2P_{5}')}{4 (-1+P_{5})}) e^{\int (\frac{ ((\mp \sqrt{-8b})(-1+P_{5})^{2}+\frac{\omega}{\hbar}x^{2}(P_{5}^2-2P_{5}'))}{2 xP_{5}(-1+P_{5})}+\frac{1}{2x})dx}, 
\end{equation}
with the following energy levels
\begin{equation}
E_{\phi_{1,2}}=\pm \hbar\omega(\sqrt{\frac{a}{2}}-1),\quad E_{\phi_{3,4}}=\hbar\omega(-c-\frac{3}{2}\mp \sqrt{\frac{-b}{2}}).
\end{equation}
Thus, the system can have one, two, three or four infinite sequences of levels and finite sequence of levels (singlet, doublet or triplet states) depending of the parameters ($a$,$b$,$c$ and $\omega$). A more detailed study of these cases should be performed and we let it to a future article. The fifth Painlev\'e transcendent also allows rational solutions [50] depending of the parameters.   
\section{Superintegrable systems}
Using results from Section 2, we can construct an infinite family of two-dimensional superintegrable systems (with $n_{1}\omega_{1}=n_{2}\omega_{2}=\lambda$, $n_{1}$,$n_{2}$ $\in \mathbb{N}^{*}$)
\begin{equation}
H=\sum_{i}^{2}  \frac{P_{x_{i}}^{2}}{2}+ \frac{\omega_{i}^{2}}{8}(1+ \frac{4(P_{5(i)}+P_{5(i)}')^{2}-P_{5(i)}^{2}}{(P_{5(i)}-1)^{2}P_{5(i)}})x_{i}^{2}+\frac{\hbar^{2}}{x_{i}^{2}}(a_{i}-b_{i}-\frac{1}{8}+\frac{b_{i}-a_{i} P_{5(i)}^{2}}{P_{5(i)}}) -\hbar\omega_{i}(1+\frac{(1+2c_{i}P_{5(i)})}{2(P_{5(i)}-1)}),
\end{equation}
with ($P_{5(i)}=P_{5(i)}(\frac{\omega_{i}}{\hbar}x_{i}^{2}, a_{i},b_{i},c_{i},-\frac{1}{8})$ i=1,2). The integrals of motion are given by (using (28), (31), (42), (43)) 
\begin{equation}
K=\frac{1}{2\lambda}(H_{1}-H_{2}),\quad I_{-}=A_{1}^{n_{1}}A_{2}^{\dagger n_{2}},\quad I_{+}=A_{1}^{\dagger n_{1}}A_{2}^{n_{2}}.
\end{equation}
These integrals of motion are of order 2 and $4(k_{1}+k_{2})$. The structure function of the polynomial algebra (of order $4(k_{1}+k_{2})-1$) is given by
\begin{equation}
\Phi(N,u,H)=\prod_{i=1}^{n_{1}}Q_{1}(\frac{H}{2}+n_{1}\omega_{1}(N+u)-(n_{1}-i)\omega_{1})\prod_{j=1}^{n_{2}}Q_{2}(\frac{H}{2}-n_{2}\omega_{2}(N+u)+j\omega_{2}),
\end{equation}
\begin{equation}
Q_{i}(H_{i})=16 \prod_{k=1}^{4}(H_{i}-E_{\psi_{k}}^{(i)}),\quad E_{\psi_{1,2}}^{(i)}=\mp\hbar\omega_{i}\sqrt{\frac{a_{i}}{2}},\quad E_{\psi_{3,4}}^{(i)}=\hbar\omega_{i}(-c_{i}-\frac{1}{2}\mp \sqrt{\frac{-b_{i}}{2}}).
\end{equation}
\section{Conclusion}
We presented a new family of superintegrable systems with higher integrals of motion from results obtained in context of supersymmetric quantum mechanics [44] using a method discussed Ref.10. The integrals of motion were constructed from the fourth order ladder operators that are product of second order supercharges. We presented also the polynomial algebra generated by these integrals of motion. The classical case of systems allowing fourth order ladder operators would differ of the quantum case. We let to a future article to obtain and study such systems and the corresponding classical superintegrable system. 

The relation between superintegrability, ladder operators and supersymmetric quantum mechanics can be used to study known superintegrable systems or obtain new superintegrable systems with higher order integrals of motion [10,13,39] and need to be more explored. 

The general equations for third order interwining relations for stationary Schrödinger operators [51] and second order intertwining relations for non-stationary Schrödinger operators [52] were obtained and may be used to construct superintegrable systems.

Moreover, systems written in terms of the Painlev\'e transcendents appear in many contexts in quantum mechanics [6-11,13,14,52-58] and recently a superintegrable system involving the sixth Painlev\'e transcendent was obtained [14]. The results obtained in this paper also extend the number of known superintegrable systems involving the Painlev\'e transcendents.
\ack The research of I.M. was supported by a postdoctoral
research fellowship from FQRNT of Quebec.

\end{document}